\documentclass[conference]{IEEEtran}
\IEEEoverridecommandlockouts
\usepackage{cite}
\usepackage{amsmath,amssymb,amsfonts,amsthm}
\usepackage{algorithmic}
\usepackage{graphicx}
\usepackage{textcomp}
\usepackage{caption}
\usepackage{subcaption}
\usepackage{balance}
\usepackage{enumitem}
\usepackage{comment}
\usepackage{booktabs}
\usepackage{url}
\usepackage{hyperref}
\usepackage[normalem]{ulem}
\usepackage{array}
\usepackage{color}
\usepackage{breqn}

\usepackage{multirow}
\usepackage{multicol}

\usepackage[]{algorithm2e}
\usepackage{xcolor}
\def\BibTeX{{\rm B\kern-.05em{\sc i\kern-.025em b}\kern-.08em
    T\kern-.1667em\lower.7ex\hbox{E}\kern-.125emX}}

\newcommand{\N}{\mathcal{N}}
\newcommand{\E}{\mathcal{E}}
\newcommand{\G}{\mathcal{G}}

\renewcommand{\footnoterule}{%
  \kern -3pt
  \hrule width 0.49 \textwidth height 0.5pt
  \kern 1pt
}
\usepackage[absolute]{textpos}
\TPMargin{5pt}

\begin{document}

\title{Distributed Optimization in Distribution Systems with Grid-Forming and Grid-Supporting Inverters\\

\thanks{This work was partially supported by NSF Career award no. 1944142 and U.S. Department of Energy under Contract DE-AC05-76RL01830.}
}
\vspace{-3pt}
\author{\IEEEauthorblockN{Rabayet Sadnan and Anamika Dubey}\\
\vspace{-10pt}
\IEEEauthorblockA{{School of Electrical Engineering and Computer Science} \\
{Washington State University} \\
}
\vspace{-11pt}}

\maketitle
\begin{abstract}
With massive penetrations of active grid-edge technologies, distributed computing and optimization paradigm has gained significant attention to solve distribution-level optimal power flow (OPF) problems. However, the application of generic distributed optimization techniques to OPF problems leads to a very large number of macro-iterations or communication rounds among the distributed computing agents delaying the decision-making process or resulting in suboptimal solutions. Moreover, the existing distribution-level OPF problems typically model inverter-interfaced distributed energy resources (DERs) as grid-following inverters; grid-supporting and grid-forming functionalities have not been explicitly considered. The added complexities introduced by different inverter models require further attention to developing an appropriate model for new types of inverter-based DERs and computationally-tractable OPF algorithms. In this paper, we expand the distribution-level OPF model to include a combination of the grid-forming, grid-supporting, grid-following inverter-based DERs and also present the application of a domain-specific problem decomposition and distributed algorithm for the topologically radial power distribution systems to efficiently solve distribution-level OPF problem. 
\end{abstract}


\section{Introduction}
Massive integration of distributed energy resources (DERs) in the power distribution systems requires applications of optimal power flow (OPF) methods to coordinate their operations \cite{molzahn2017survey,khushalani2007development}. Although both centralized and distributed optimization techniques have been used to solve the distribution-level OPF problem, lately, the distributed optimization methods have gained significant attention to solving OPF due to their robustness to single-point failures \cite{molzahn2017survey}. In addition, the scalability and complexity of the centralized OPF (C-OPF) mechanism, which stems from both the size of the network and the increased inverter-interfaced DERs, can be managed using distributed optimization methods \cite{molzahn2017survey}. However, the direct application of the existing distributed optimization algorithms results in slow converging algorithms for OPF. Moreover, the DERs are generally coupled with smart inverters, with possible grid-following and grid-supporting functionalities, capable of providing grid services such as voltage and frequency support \cite{palmintier2016feeder,epri_SI}. Incorporating different inverter operating modes can further increase the computational complexities and make it more challenging for state-of-art D-OPF algorithms to converge within a reasonable time or number of macro-iterations.

Typically, the models of grid-forming \& supporting inverters are either developed for dynamic simulations of distribution systems or modeled for microgrid operations \cite{du2020modeling,rocabert2012control,vergara2018generalized}. Although in \cite{khushalani2007development,garcia2004improvements}, such inverter-connected buses are modeled for the quasi-static power-flow study of grid-connected distribution systems, the appropriate models to formulate the OPF, especially in distributed optimization setting, require further attention. A majority of the existing OPF literature, the inverter-interfaced DERs are usually modeled as negative loads assuming a grid-following functionality \cite{molzahn2017survey,zheng2015fully,dall2013distributed}. Upon solving the OPF, the grid-following DERs are set to dispatch the optimal active and/or reactive power to the network. In \cite{du2020modeling,rocabert2012control}, only dynamic simulation cases have been considered while developing the models for grid-forming and supporting inverters; thus, they can not be directly adopted in OPF formulations. In \cite{vergara2018generalized}, authors developed a generalized model of such inverters for quasi-static OPF problems; however, they assumed the inverters would operate exclusively in a grid-following mode in the grid-connected setting. Recent work models the grid-supporting inverts in the OPF formulation by adding the Q-V droop constraints to the formulation; however, they solve a convex relaxed problem centrally to reduce the resulting compute complexities \cite{savasci2021distribution}. Note that depending upon their settings, DERs can operate in grid-forming and grid-supporting mode in the grid-connected distribution networks \cite{garcia2004improvements,khushalani2007development, epri_SI}. To the authors' best knowledge, the existing work does not include a comprehensive model of different operational modes of inverter-interfaced DERs quasi-static OPF problems for power distribution systems.

In addition, the state-of-the-art distributed optimization methods for OPF, such as Alternating Direction Method of Multipliers (ADMM), Auxiliary Problem Principle (APP), primarily suffer from a large number of required communication rounds among distributed agents to solve one instance of the problem \cite{zheng2015fully,peng2016distributed}. A large number of communication rounds for solving one time-step of the OPF problem is not desirable in power distribution systems, as it will lead to crucial delays in the decision-making process. Further, the intermediate iterates can fail to satisfy the power flow equations leading to violation of the critical power systems operating constraints \cite{molzahn2017survey}. This problem of slow convergence will get aggravated upon including more complex grid-edge devices such as inverter-based DERs with grid-supporting and grid-forming modes of operation.

To address these limitations, previously, we have proposed a novel distributed OPF (D-OPF) algorithm that uses specialized problem decomposition and information exchange protocols. The proposed approach actively leverages the topologically radial distribution feeder; it significantly reduces the communication rounds (by order of magnitude) needed to converge compared to ADMM based D-OPF methods \cite{rabayet_Dist}. However, our prior work does not include different modes of inverter operation; we only considered grid-following inverters in the D-OPF formulation. This paper aims to model different modes of operations for inverter-interfaced DERs in the quasi-static OPF problem formulation and demonstrate the applicability of the previously developed D-OPF algorithm to solve OPF problems for distribution systems with a combination of grid-forming, grid-following, and grid-supporting inverter-interfaced DERs. Specifically, we develop the models for grid-following, grid-supporting, and grid-forming DER inverter in the distribution-level OPF formulation. Next, we evaluate the applicability of a faster and scalable D-OPF algorithm for a combination of grid-forming and grid-supporting inverters (with various penetration levels). The solution quality and the performance of the D-OPF algorithms are compared with the state-of-the-art ADMM based D-OPF method. 


\section{Power Flow \& DER Models}
In this paper, $(\cdot)^T$ represents matrix transpose; $|~.~|$ symbolizes the absolute value of a number or the cardinality for a discrete set; $(\cdot)^{(n)}$ represents the $n^{th}$ macro-iteration;
In this section, first, we discuss about the network model, and then we detail the DER models. The DERs are modeled as photovoltaic modules interfaced using smart inverters, capable of four-quadrant operation. These inverters can be (i) grid following (GFLI), (ii) grid supporting (GSI), or (iii) grid forming (GFI) DERs \cite{rocabert2012control}. Please note, traditional DGs can also be incorporated in the model.

\subsection{Nonlinear Network Model}
Let us assume a balanced, radial power distribution network, represented by the directed graph $\G = (\N, \E)$, where $\N$ and $\E$ be the set of all nodes $j$ and all distribution lines connecting the ordered pair of buses $\{ij\}$ in the system. Let $r_{ij}$ \& $x_{ij}$ be the series resistance \& reactance $\forall\{ij\}\in \E$. In $k: j\rightarrow k$, $k$ represents the children nodes for the node $j$. We denote $v_j$ and $l_{ij}$ as the squared magnitude of voltage (at node $j$) and current flow (in branch $\{ij\}$), respectively. Also, complex power $S_{L_j} = p_{L_j}+jq_{L_j}$ is the load connected and $S_{D_j} = p_{Dj}+jq_{Dj}$ is the output power of DER and $ q_{Cj}$ is the capacitor at node $j$. The network is modeled using the branch flow equations \cite{baran1989optimal1} defined for each line $\{ij\}\in \E$ and $\forall j \in \N$ in \eqref{pfmodel}. 

 \vspace{-0.3cm}
\begin{small}
\begin{IEEEeqnarray}{C C}
\small
\IEEEyesnumber\label{pfmodel} \IEEEyessubnumber*
P_{ij}-r_{ij}l_{ij}-p_{L_j}+p_{Dj}= \sum_{k:j \rightarrow k} P_{jk}   \label{pfmodel1}\\
Q_{ij}-x_{ij}l_{ij}-q_{L_j}+ q_{Cj}+q_{Dj}= \sum_{k:j \rightarrow k} Q_{jk} \label{pfmodel2}\\
v_j=v_i-2(r_{ij}P_{ij}+x_{ij}Q_{ij})+(r_{ij}^2+x_{ij}^2)l_{ij}\label{pfmodel3}\\
v_il_{ij} = P_{ij}^2+Q_{ij}^2 \label{pfmodel4}
\end{IEEEeqnarray}
\end{small}

\subsection{Grid Following Inverter-interfaced DERs}
Generally in OPF for power distribution networks, DERs are modeled as grid following inverters -- that can generate power within their physical limits for optimal operations. A negative load model is adopted to model them in OPFs, i.e., $q_{Dj}$ or $p_{Dj}$ or both is considered as decision variables \cite{molzahn2017survey}. In the grid following mode, if the reactive power generation, $q_{Dj}$, is modeled as the decision variable for the optimal operation, then the real power generation by the DER, $p_{Dj}$, is assumed to be known (measured). Let the rating of the DER connected at node $j$ be $S_{DRj}$, then the limits on $q_{Dj}$ are given by \eqref{DG_lim}. On the contrary, if the active power generation, $p_{Dj}$, is modeled as the decision variable, then $q_{Dj}$ is set to $0$, and $p_{Dj}$ can vary between $0$ and $S_{DRj}$, see \eqref{DG_lim2}.

\vspace{-0.3cm}
\begin{small}
\begin{IEEEeqnarray}{C C}
\small
\IEEEyesnumber\label{PQDER} \IEEEyessubnumber*
-\sqrt{S_{DRj}^2-p_{Dj}^2} \leq q_{Dj} \leq \sqrt{S_{DRj}^2-p_{Dj}^2}   \label{DG_lim}\\
\text{Or, }\hspace{0.2 cm}0 \leq p_{Dj} \leq S_{DRj} \label{DG_lim2}
\end{IEEEeqnarray}
\end{small}

\subsection{Grid Supporting Inverter-interfaced DERs}
\begin{figure}[t]
    \centering
    \includegraphics[width=0.25\textwidth]{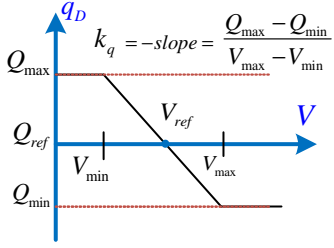}
        \vspace{-0.2cm}
    \caption{Droop Curve for GSIs}
    \label{droop_curve}
\vspace{-0.5cm}
\end{figure}

The GSI DERs deliver proper active and reactive power to contribute to the grid operations, such as, frequency and voltage. The DERs interfaced with GSIs are represented either as an current-source based or voltage-source based converters \cite{rocabert2012control}. Generally, the optimal actions from these DERs are extracted by implementing droop curves, and dynamically modify their operating points. However, the frequency of changing the droop operating points might be lower than the frequency of solving OPFs due to the communication constraints associated with GSIs. Thus, it is often requires to optimize considering a droop curve of the GSI, rather than optimize the droop curve itself \cite{savasci2021distribution,epri_SI}. Please note, although the current-source and voltage-source based GSIs have different control loops in the hardware, but they have similar droop characteristics with negative slopes \cite{epri_SI,rocabert2012control}; thus, their model for OPF formulation is same. Here, to model the GSIs, Q-V droop curves as depicted in Fig. \ref{droop_curve} are considered instead of 4 break-point piece-wise linear curves (see \cite{epri_SI}: Fig. 11-1), as the later ones are usually configured optimally \cite{epri_SI}. 

\vspace{-0.3cm}
\begin{small}
\begin{equation}
q_D (V) = Q_{ref} + k_q (V_{ref}-V)   \label{droop1}
\end{equation}
\end{small}

The Q-V droop curve for the GSIs are detailed in \eqref{droop1}; where the reactive power output, $q_D$ is dependent on the nodal voltage V, and $k_q$ is the negative of the slope of the droop curve. An example of this linear relation, $q_D(V)$, is depicted in Fig. \ref{droop_curve}. However, the network model does not use the magnitude V, rather use the squared of that value, $v$. So, to be able to use the droop curves in the OPF formulation, the $q_D(v)$ curve has to be formulated.

Mostly in power distribution systems, the nodal voltage is maintained between $0.95$ to $1.05$ pu. For $V_{ref} \in [0.95, 1.05]$, then we can approximate the $q_D(V)$ relation by using Taylor's series. Around the reference voltage, $V_{ref}$, voltage $v$ can be approximated by \eqref{T_app}. Using this approximation, we get the $ q_{Dj}(v)$ relation in \eqref{droop2} for a GSI node $j$. 
The equation \eqref{droop2} can be used for droop curves with any slope and reference points.

\vspace{-0.3cm}
\begin{small}
\begin{IEEEeqnarray}{C C C}
\small
\IEEEyesnumber\label{T_app} \IEEEyessubnumber*
v = V_{ref}^2 + 2\times V_{ref}(V -V_{ref})  \label{T_app1}\\
\Rightarrow V_{ref}-V = \frac{V_{ref}^2-v}{2V_{ref}} \label{T_app21}
\end{IEEEeqnarray}
\end{small}
\vspace{-0.3cm}
\begin{small}
\begin{equation}
 q_{Dj}(v_j) = Q_{ref,j} + k_{q,j} \Big(\frac{V_{ref,j}^2-v_j}{2V_{ref,j}}\Big)   \label{droop2}\\
\end{equation}
\end{small}
\vspace{-0.35cm}

\subsection{Grid Forming Inverter-interfaced DERs}
The DERs with grid forming capabilities act as an ideal voltage source that can generate a specified voltage and maintain the  system's frequency. Specifically, they provide the voltage and frequency support in case islands are formed. In grid-connected mode, they still provide a firm voltage at the point of common coupling \cite{khushalani2007development}. To solve the quasi-static OPF problem in grid-connected mode for GFIs, the voltage of such nodes are kept fixed, while the real and reactive power generation is modeled as the decision variable for that node. The model for grid-connected GFIs are detailed in \eqref{GFI_model}. Here, for the GFI at node $j$, the voltage is set to $v_{j,set}$ \eqref{GFI_model1}; the real power generation $p_{Dj}$, and the reactive power generation $q_{Dj}$ is limited by the physical limit of that DER, $S_{DRj}$ \eqref{GFI_model2}.

\vspace{-0.3cm}
\begin{small}
\begin{IEEEeqnarray}{C C}
\small
\IEEEyesnumber\label{GFI_model} \IEEEyessubnumber*
v_j = v_{j,set} \label{GFI_model1}\\
p_{Dj}^2+ q_{Dj}^2\le S_{DRj}^2 \label{GFI_model2}
\end{IEEEeqnarray}
\end{small}
\vspace{-0.3cm}

In GFI models, the real power generation, $p_{Dj}$, is considered as a decision variable at node $j$. Thus, a stable quasi-static operation point can be achieved, and the angle stability is ensured. As the power flow solutions are unique for radial networks \cite{Baran_uniq}, fixing multiple nodal voltages with fixed real power generations at those nodes can lead to angle instability -- causing unavailability of feasible power flow solutions.

\section{Optimal Power Flow Problem Formulation}
In this section, first, the centralized optimal power flow formulation is developed. Then the decomposition approach and the distributed optimization problem formulation is detailed.

\subsection{Centralized OPF (C-OPF) Problems}
In this section a centralized OPF problem is formulated for power distribution system with different types of DERs. The problem is defined by a network-level problem objective, the power flow models in \eqref{pfmodel}, and the operating constraints on the power flow variables. In this paper, we formulate the active power loss minimization problem; the problem objective is to reduce the network losses by controlling the power output from DERs. Let $X=[P_{ij}, ~Q_{ij}, ~l_{ij}, ~v_j, ~p_{Dj}, ~q_{Dj}]^T$ be the problem variables $\forall j\in \N$, and $\forall \{ij\}\in \E$. Note that, if $p_{Dj}$, $q_{Dj}$ is known and uncontrollable, then we set these values at node $j$ with their known measurements. Also, let $F (X )$ be the cost function representing the total power loss in the given distribution system. Then, the OPF problem is defined as the following in \textbf{(C1)}. Here, $\N^{GSI}$ and $\N^{GFI}$ denotes the sets of GSI and GFI buses, respectively;  ${V_{min}} = 0.95$ and ${V_{max}} = 1.05$ are the limits on bus voltages, and $(I^{rated}_{ij})^2$ is the thermal limit for the branch $\{ij\}$.

\vspace{-0.4cm} 
\begin{small}
\begin{IEEEeqnarray}{C C}
\IEEEyesnumber\label{LM_COPF} \IEEEyessubnumber*
\text{\textbf{(C1)}}\hspace{0.4cm} \min \hspace{0.2cm} F (X ) = \sum_{\{ij\}\in \E} l_{ij}r_{ij} \hspace{0.6cm}\\
\text{s.t.}  \hspace{0.2cm} \text{\eqref{pfmodel}, \eqref{PQDER}; \eqref{droop2}}\hspace{0.1cm}\forall j\in \N^{GSI}\text{, and \eqref{GFI_model}}\hspace{0.1cm}\forall j\in \N^{GFI} \label{C11}\\
{V_{min}^2} \leq v_j \leq {V_{max}^2} \hspace{0.7cm} ;\forall j\in \N \label{LM_OPF1}\\
l_{ij} \leq \left(I^{rated}_{ij}\right)^2  \hspace{0.5cm} ;\forall \{ij\} \in \E  \label{C12}
\end{IEEEeqnarray}
\end{small}
\vspace{-0.4cm}

\subsection{Distributed OPF (D-OPF) Problems}
The OPF problem described by \textbf{(C1)} can be solved by several decision-making agents in parallel, and over macro-iterations, they can get a consensus in the shared variables to achieve overall convergence. For the distributed OPF, the decomposition approach is adopted from \cite{rabayet_Dist}, that leverages the radial topology of the power distribution networks. The network is assumed to be decomposed into several areas, each with one decision-making agent. Thus, the C-OPF problem is decomposed into several sub-problems, where each sub-problem is associated with one area. While solving the local sub-problem, the shared boundary variables with the upstream area (UA) and the downstream areas (DAs) are approximated as a fixed voltage source and fixed loads, respectively. These fixed values are changed at every macro-iteration step and set equal to the neighbors' computed value of that respective shared bus variable, from the previous macro-iteration step.

Suppose the network is composed of $N$ areas- $\{A_1,A_2,$ $ A_3, ...,$ $A_{N}\}$, and $A_m=$ $\G(\N_m,\E_m)$ for $m =$ $\{1,2,..., N\}$. Let $X_m$ be the set of local variables and $X = \bigcup_{m=1}^{N} X_m$. Also, area $A_m$ shares bus $0$ with its UA and node $k$ (where $\{jk\} \in \E_m$) is shared with the DA. Since each shared bus is solved by both neighboring areas, a subscript notation has been introduced for those shared nodes -- the subscript of a node represents the area, that has solved the variable; e.g., $v_{0_m}$, $p_{k_m}$ are solved by area $A_m$, where $v_{0_{ua}}$, $P_{jk_{da}}$ are solved by UA and DA, respectively. At each $A_m$ area, the optimization problem \textbf{(D1)} is solved $\forall \{ij\}\in \E_m$ and $\forall j \in \N_m$ at macro-iteration step $n$. Constraint \eqref{eqv1} represents the UA \& DA approximations. The macro-iteration stops when the shared boundary variables reach a consensus among all the neighbors, and then the decision variables are dispatched within the area. The consensus at the boundary can also be achieved using other Fixed Point Iteration methods \cite{rabayet_Dist,sadnan2021online}.


\vspace{-0.4cm}
\begin{small}
\begin{IEEEeqnarray}{C C}
\IEEEyesnumber\label{lin_D_OPF} \IEEEyessubnumber*
\text{\textbf{(D1)}}\hspace{0.4cm} \min \hspace{0.2cm} F (X_m^{(n)}) = \sum_{\{ij\}\in \E_m} l_{ij}^{(n)}r_{ij} \hspace{0.6cm}\\
\text{s.t. \eqref{C11}-\eqref{C12}}\\
v_{0_m}^{(n)} = v_{0_{ua}}^{(n-1)}\label{eqv1}; \hspace{0.1cm}
p_{k_m}^{(n)} = P_{jk_{da}}^{(n-1)}; \hspace{0.1cm} 
q_{k_m}^{(n)} = Q_{jk_{da}}^{(n-1)} 
\end{IEEEeqnarray}
\end{small}
\vspace{-0.6cm}


\section{Numerical Simulations \& Results}
This section presents a detailed evaluation of the D-OPF algorithm with GFLI, GSI, and GFI DERs. Three test cases have been simulated: (a) Case I: All the DERs are in grid following mode, (b) Case II: Various levels of DERs in grid supporting mode, and (c) Case III: GFLI, GFI and GSI inverters are active in the network with some synchronous DG buses -- considered as generator bus (PV type bus).

\vspace{-0.1cm}
\subsection{Simulation Setup}
\begin{figure}[t]
    \centering
    \includegraphics[width=0.42\textwidth]{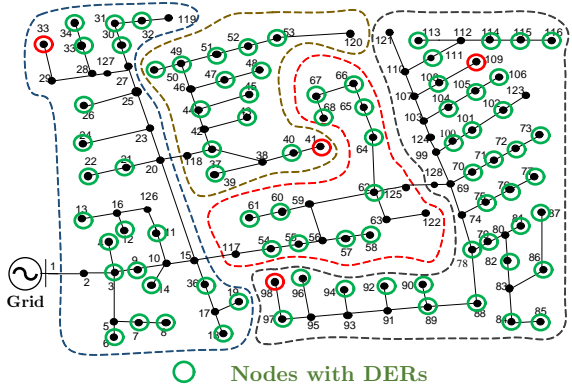}
        \vspace{-0.2cm}
    \caption{IEEE 123-bus Test System: Composed with 4 Areas}
    \label{123_bus}
\end{figure}

The balanced IEEE 123-bus test system, assumed to be composed of four areas, is used for numerical studies (Fig. \ref{123_bus}). Each Area has a local compute agent, that solves local sub-problems and communicates with the other neighboring agents. 85 DERs are placed at every load nodes. The nominal power rating of all the DERs is 42\% of the load at that bus. For GSIs, the droop curve depicted in Fig. \ref{droop_curve} is adopted. The substation voltage is set to $1.03$ pu, and the solution time for D-OPF is defined as the summation of the maximum time required by agents at each macro-iteration step.

\begin{figure}[t]
\hspace{-.3cm}
\centering
\begin{subfigure}{.24\textwidth}
  \centering
  \includegraphics[width=1.1\linewidth]{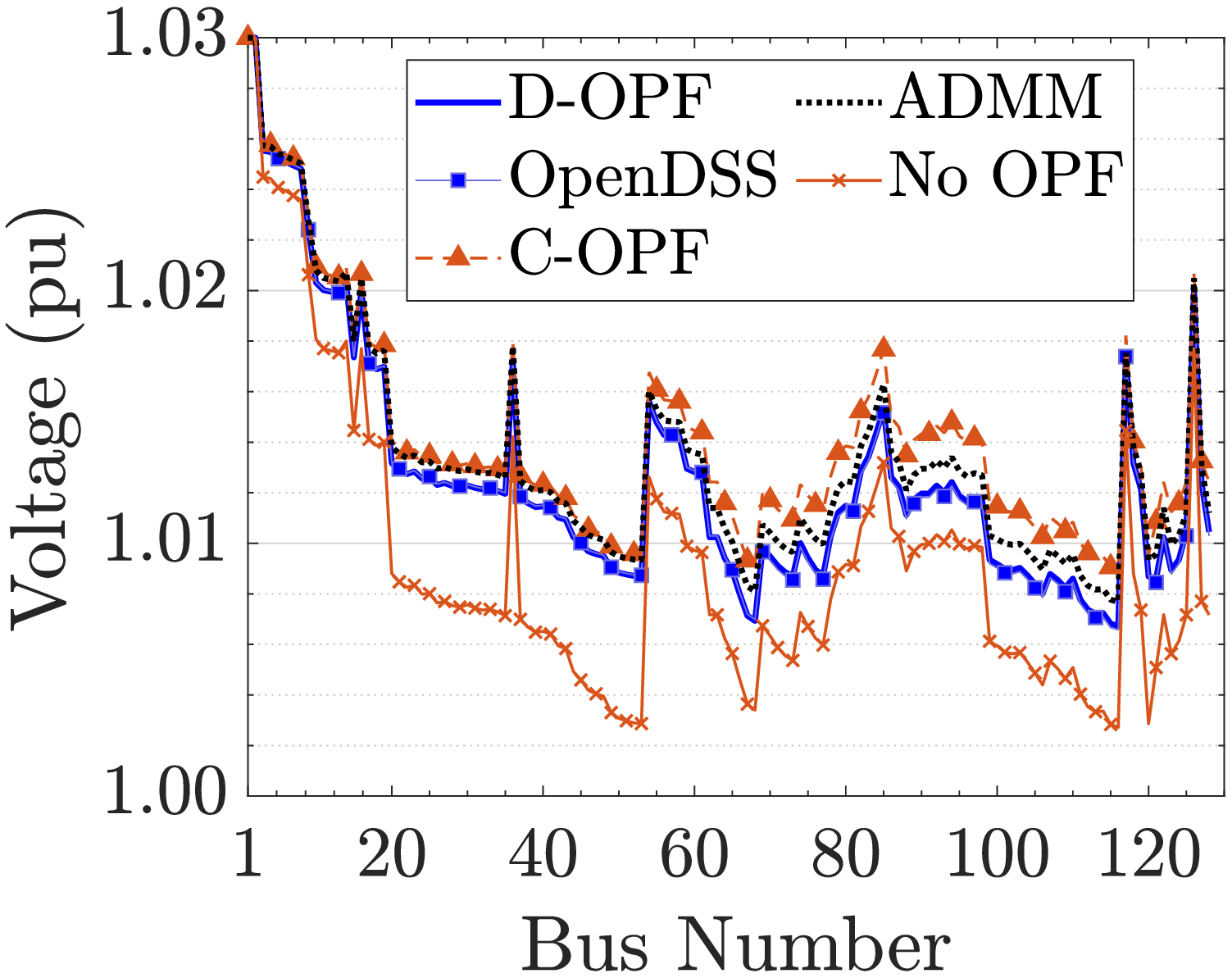}
  \caption{Nodal voltages}
  \label{volt_GFLI}
\end{subfigure}
\hspace{-0.1cm}
\begin{subfigure}{.24\textwidth}
  \centering
  \includegraphics[width=1.1\linewidth]{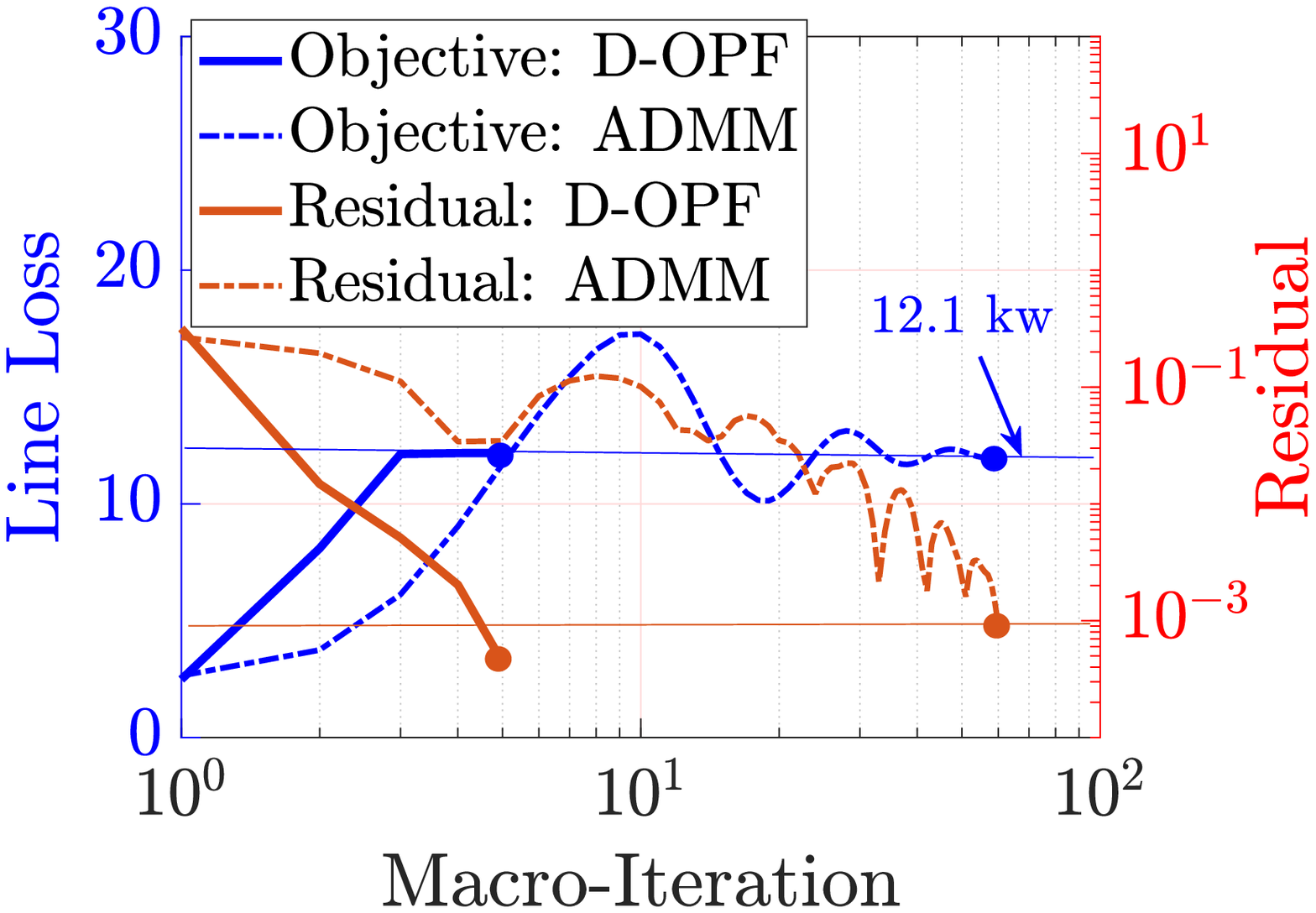}
  \caption{Convergence}
  \label{R_GFLI}
\end{subfigure}
\caption{\centering{Case I: Numerical Results }}
\vspace{-0.6cm}
\label{Result_GFLI}
\end{figure}

\subsection{Case I: Validation and Comparison}
In this case, all the DERs are assumed to be operating in grid following mode, i.e., generating the specified P and Q. This case validates the effectiveness of the proposed D-OPF over state-of-the-art ADMM based methods, such as \cite{zheng2015fully}. From the Table \ref{res_tab1}, we can see the line loss for the D-OPF is 12.18 kW, which is very close to the central solution of 12.10 kW, and it only takes 4 macro-iterations to reach a consensus. The nodal voltages of D-OPF and C-OPF have also been compared in Fig. \ref{volt_GFLI}. Additionally, the D-OPF solutions have been validated using OpenDSS (see Fig. \ref{volt_GFLI}) -- the decision variables from the  D-OPF solution is implemented in OpenDSS. Further, the ADMM based distributed algorithm has been compared; though the solutions are similar -- compared with central or simulated D-OPF method, the number of macro-iterations is more than 10 times higher (60 iterations) than the proposed D-OPF method (4 iterations).

\begin{figure}[t!]
    \centering
    \includegraphics[width=0.42\textwidth]{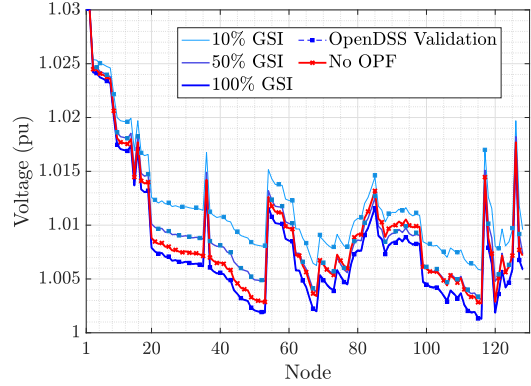}
        \vspace{-0.2cm}
    \caption{Case II: Voltages for Different Levels of GSIs}
    \label{V_GSI_comp}
\end{figure}

\begin{small}
\begin{table}
    \vspace{-0.3cm}
    \centering
    \caption{Comparison of OPF Solutions}
    \vspace{-0.1cm}
    \label{res_tab1}
    \setlength{\tabcolsep}{2.2pt}
    \begin{tabular}{|c|c| c| c|  c|  c|}
    \hline
    \multicolumn{2}{|c|}{\textbf{Test Cases}} &{} & \textbf{C-OPF}& \textbf{D-OPF}  & \textbf{Iteration}\\
    \hline
    \multirow{2}{*}{\textbf{Case I}}&\multirow{2}{*}{100\% GFLI} & Loss & 12.10 kW  & 12.18 kW  &  \multirow{2}{*}{4}\\\cline{3-5}
    & & Time & 14.93 sec & 13 sec &\\
    \hline
    \multirow{6}{*}{\textbf{Case II}}&\multirow{2}{*}{10\% GSI} & Loss & 12.6 kW  & 12.4 kW  &  \multirow{2}{*}{4}\\\cline{3-5}
    & & Time & 12.6 sec & 12.13 sec &\\\cline{2-6}
    &\multirow{2}{*}{50\% GSI} & Loss & 12.9 kW  & 13.0 kW  &  \multirow{2}{*}{6}\\\cline{3-5}
    & & Time & 8.15 sec & 11.5 sec &\\\cline{2-6}
    &\multirow{2}{*}{100\% GSI} & Loss & 13.9 kW  & 13.9 kW  &  \multirow{2}{*}{8}\\\cline{3-5}
    & & Time & 2.15 sec & 2.2 sec &\\
    \hline
        \multirow{2}{*}{\textbf{Case III}}&\multirow{1}{*}{80\% GFLI, 5\% GFI, 5\% } & Loss & 9.6 kW  & 9.87 kW  &  \multirow{2}{*}{36}\\\cline{3-5}
    {}& \multirow{1}{*}{PV type and 10\% GSI} & Time & 30 sec & 100 sec &{} \\
    \hline
    \end{tabular}
    \vspace{-0.3cm}
\end{table}
\end{small}

\subsection{Case II: Impacts of GSI DERs}
In this case, various numbers of GSI DERs have been introduced in the system to study the impacts of such converters in the power distribution networks. Also, the performance of the simulated D-OPF have been investigated. Three different levels -- $10\%$, $50\%$, $100\%$, of the DERs are assumed to be with GSI functionalities for this case. For $10\%$ GSIs, nodes 7, 22, 39, 48, 58, 60, 90 and 105 are selected (Fig. \ref{123_bus}). For $50\%$ scenario, 42 DERs are selected randomly from those 85 DERs. From the Table \ref{res_tab1}, it is observed that with increasing GSIs, the line losses in the system increases -- 12.4 kW to 13.9 kW. Generally for loss minimization, the Q dispatches and the voltages are gravitated towards the max value, however, the droop curves dictates the Q dispatches based on the voltages -- activating more constraints on the Q dispatches, causing higher losses with increased GSIs. On the other hand, though the macro-iterations slightly increases (4 to 8), the overall solution time decreases (12 to 2 sec) as the solution space is reduced by the added constraints. For ADMM based algorithm, it takes 60-70 macro-iterations for these cases. The nodal voltages for these scenarios are also compared and validated using OpenDSS. The D-OPF and OpenDSS voltages are exactly same -- validating the solution quality of D-OPF with GSIs (Fig. \ref{V_GSI_comp}). Also, with increased GSIs, the voltages are decreased -- causing higher line losses.

\subsection{Case III: Impacts of GFI DERs}
\begin{figure}[t]
\hspace{-.3cm}
\centering
\begin{subfigure}{.24\textwidth}
  \centering
  \includegraphics[width=1.1\linewidth]{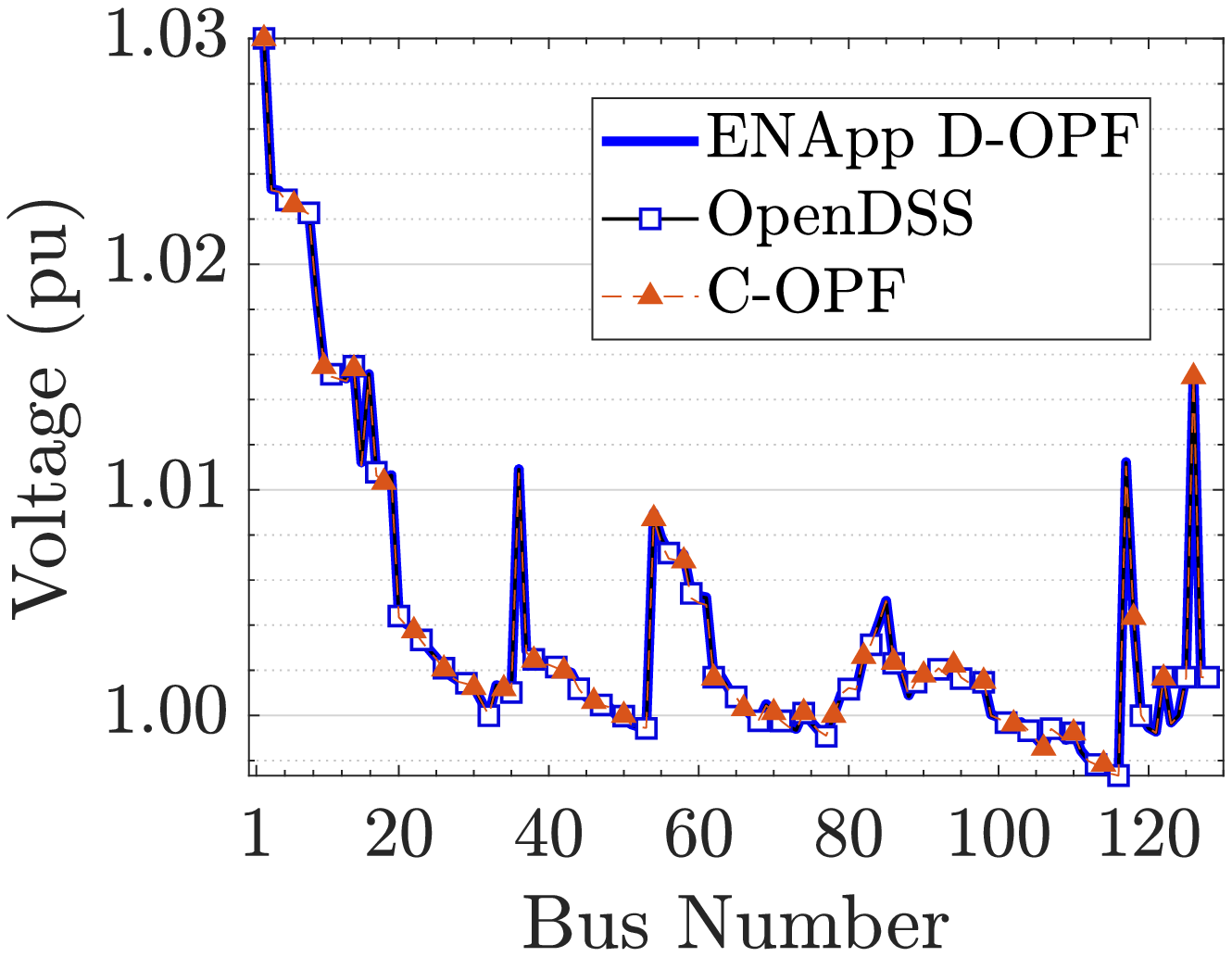}
  \caption{Nodal voltages}
  \label{volt_GFI}
\end{subfigure}
\hspace{-.2cm}
\begin{subfigure}{.24\textwidth}
  \centering
  \includegraphics[width=1.1\linewidth]{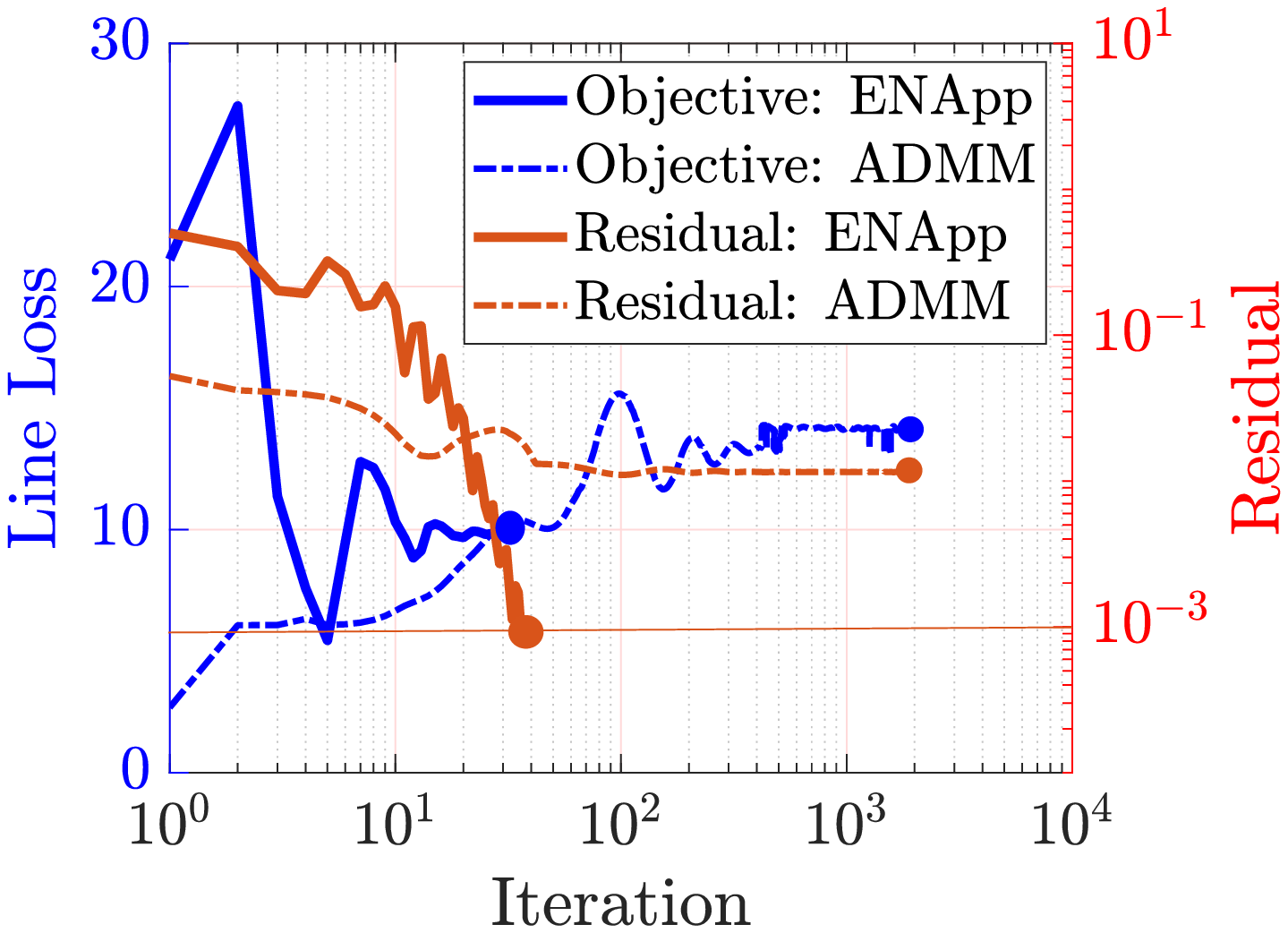}
  \caption{Convergence}
  \label{R_GFI}
\end{subfigure}
\caption{\centering{Case III: Numerical Results}}
\label{Result_GFI}
\end{figure}

The last case implements DERs -- interfaced with GFLI, GSI and GFI, to show the solution quality and the robustness of the simulated D-OPF. While $10\%$ nodes are GSIs, nodes 32, 50, 67 \& 78 are assumed to be GFI buses. The rating of the GFI DERs are increased 10 times to maintain the specified voltage at those nodes. We also included some DGs (node 33, 41, 98 and 109), that can operate as a generator bus (PV type bus) as mentioned in \cite{khushalani2007development}. We include an added penalty term with the cost function ($F(X) = \sum l_{ij}r_{ij}+ M\sum_{j\in \N^{PVBus}}(v_j - v_{set})$) to model such generator buses in our OPF. Here, $M$ is a high number ($10^2$) and $v_{set}$ is the specified voltage at that node. Within the physical limits, these buses tries to maintain the voltage and dispatches the set P value \cite{khushalani2007development}. All the specified voltages for this case are set to $1.00$ pu. The inclusion of GFIs increases the macro-iteration number to 36 for D-OPF (Table \ref{res_tab1}), but the solution matches with the central OPF; the result is also validated against OpenDSS (Fig. \ref{volt_GFI}). On the contrary, the ADMM based method takes more than 2000 macro-iterations to converge, and the solution gives a sub-optimal result (Fig. \ref{R_GFI}). This showcases that the inclusion of GFI DERs can increase the computational time, however the simulated D-OPF method is robust enough to reach the global solution, where the ADMM based method fails to reach that -- even with 2000 iterations. 

\section{Conclusion}
We developed models for the inverter-interfaced DERs operating in grid-forming and grid-supporting modes for the quasi-static OPF problems in the power distribution system. A specialized distributed optimization algorithm that actively leverages the distribution system's radial topology (in structure) is used to solve the resulting OPF problem in a distributed manner. Compared to ADMM, the simulated distributed OPF (D-OPF) algorithm not only reduced the number of macro-iterations required to converge by orders of magnitude but also converged to the same solution as the centralized OPF method. Note that for some cases, especially with a large number of grid-forming inverters, ADMM-based distributed OPF converged to a higher cost even after thousands of macro-iterations. This paper is the first to incorporate models for grid-forming and grid-supporting functions of DERs (along with grid-following) in the OPF problem and demonstrate the use of a computationally tractable and scalable D-OPF algorithm to solve the resulting more complex OPF problem efficiently. This is also the first study to demonstrate that the specialized D-OPF algorithm outperforms state-of-the-art ADMM-based D-OPF algorithms for radial distribution feeders connected to various inverter-based DER technologies.

\balance
\bibliographystyle{IEEEtran}
\bibliography{PESGM_2022}

\end{document}